# PARSIMONIOUS DATA

## HOW A SINGLE FACEBOOK LIKE PREDICTS VOTING BEHAVIOUR IN MULTIPARTY SYSTEMS

Kristensen Jakob Baek[a,1], Albrechtsen Thomas[b], Dahlgaard Emil[b], Jensen Michael[b], Skovrind Magnus[c], Bornakke Tobias[d]

Category of submission: Social, Political Sciences

[a] School of Social and Political Sciences, University of Canterbury, New Zealand

[b] Department of Research, Nextwork A/S, Denmark

[c] Department of Research, Analyse & Tal F.M.B.A, Denmark

[d] Department of Sociology, University of Copenhagen, Denmark

[1] Corresponding author: Jakob Baek Kristensen

Email: jakob.kristensen@pg.canterbury.ac.nz

Address: 12F Kirkwood Avenue, 8041, Ilam, Christchurch, New Zealand

Phone: +640224328571

# Abstract

Recently, two influential PNAS papers have shown how our Facebook likes can accurately predict details about our personal traits and political attitude (1, 2). In this paper, we show how a more parsimonious measure based solely on likes directed toward posts from political actors can accurately predict present day voter intention even within a challenging multiparty setting. Combining the online and offline, we connect a subsample of surveyed respondents to their public Facebook activity and apply machine learning classifiers to explore the link between their political liking behaviour and actual voting intention. Through this work, we show how even a single selective Facebook like, can reveal as much about our political voter intention as hundreds of heterogeneous likes. Further, by including the entire political like history of the respondents, our model reaches prediction accuracies above previous multiparty studies (60-70%). On this basis, we challenge the ideal of 'bigger and broader data': while Facebook likes pertaining to everything from reality stars to soil types might accurately predict political observation, even more accurate results can be reached through contexts-specific parsimonious data strategy. The main contribution of this paper is to show how public like-activity on Facebook, through a selective data strategy allows political profiling of individual users in a multiparty system with accuracies above previous studies. Beside increased accuracies, the paper shows how such parsimonious data strategies allows us to generalize our findings to the entire population and even across national borders to other political multiparty systems.

**Significance Statement**

This study shows how liking politicians' public Facebook posts can be used as an accurate measure for predicting voter intention in a multiparty system. We highlight a few, but selective digital traces produce prediction accuracies that are on par or even greater than many current approaches based upon big and broader datasets. The approach in this study relies on data that are publicly available, and the simple setup, we propose can, with some limitations, be generalized to millions of users in countries with multiparty systems.

# Introduction

The representative opinion survey have long been the pinnacle of empirical research in political science (3, 4). The immense recent growth in digital platforms has provided researchers with the possibility to study human behaviour on a whole new scale from the traces left behind from our digital interactions (5). From being limited to surveys with a couple of thousand respondents, political studies covering millions of people have emerged with the field of computational social science generating important new knowledge about our digital and analogue lives.

Within the subfield of election forecasting, Scholars have shown the potential for predicting election outcomes based on digital data from a diverse list of platforms including Youtube (6), Google (7), Twitter (8, 9), Facebook (10, 11) and even Wikipedia (12). The big social media platforms such as Facebook and Twitter have largely been the most successful data sources, with prediction outperforming traditional pooling both in accuracy and scale (6, See 13 for a general review). While this emerging field has overly focused on predicting aggregated electoral results (14), a smaller group of studies have however focused on the task of predicting the individual political orientation of people (1, 2, 11, 15–19). Notably, (13) has shown how political profiling with Twitter data can reach very high accuracies, just as (18) convincingly has displayed political orientation can be determined by comparing peoples' writing style with the writings on politicians' public Facebook profiles. While most of these studies attain high predictions accuracies, this accuracy is often reached by using only the most active users and results are rarely validated against offline data such as questionnaires (17). This limitations have however been greatly tackled in the work of Konsinski and colleagues who, in two papers ranked among the top-10 most influential papers in the history of PNAS[*], have shown how our political attitude and other personal traits, with great accuracy, can be predicted based solely on Facebook likes (1, 2). Applying machine learning algorithms to search for patterns in the hundreds of diverse Facebook likes, these already famous experiments has thus disclosed how our preference for 'Hallo Kitty' and 'Harley Davidson' can reveal details of our personality and political attitude - often with better assessments than one's friend.

---

[*] As calculated by altmetric: https://pnas.altmetric.com/details/3058702 & https://pnas.altmetric.com/details/1294474.

The majority of studies have however zoomed in on predicting political orientation in a two-party system, hereby avoiding the more challenging task of making individual predictions in multiparty settings, our time's principal political system (18). In this paper, we make up for this limitation studying how individual party choice in a multiparty system can be predicted with high accuracy through the single measure of likes toward political actors. We base our prediction on likes toward posts from public pages of Danish parties and politicians collected between January 2015 and 2017 through the Facebook Graph API. Through machine learning based prediction models, we test how 'political likes', consisting of likes on posts by politicians and parties, are able to predict present day voter intention in a multiparty system for a subsample of surveyed respondents. The process is illustrated in Figure 1.0 while the models and results can be found in table 1.1 (details in Materials and Methods).

The main contribution of this paper is to show how public Facebook activity, even within a challenging multiparty system, can be effectively used to predict an individual's present day voter intention. Based on the simplistic measure of political likes, our models reach a prediction accuracy of between 60 and 70%, which are above any previous multiparty studies. Also, we show how even a single selective Facebook post like, can reveal as much about our present-day voter intention as hundreds of diverse likes drawn from our profile. In doing this, we wish to challenge the current trend toward broader and bigger data running through the majority of studies within computational social science. By developing the parsimonious measure of political likes, we make the point, that though likes pertaining to everything from reality stars to soil types can accurately predict personal traits, more accurate and generalizable results, when looking specifically for present day voter intention, will tend to follow from a more parsimonious data strategy.

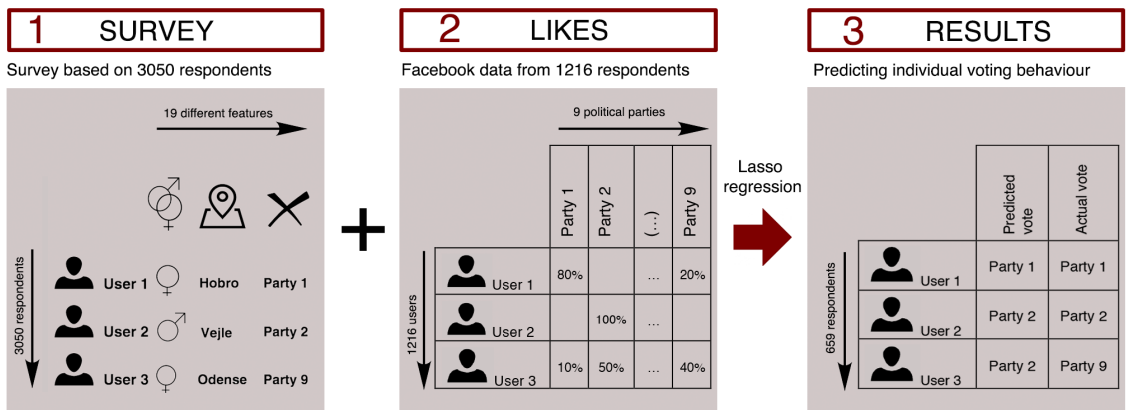

*Figure 1.0: The study progress in 3 parts: 1) A representative questionnaire was completed by 3050 randomly selected people living in Denmark, providing information on standard sociodemographic qualities, political values and present day voter intention toward parties eligible in the general election. As shown in the Table S5, the sample is overly demographically representative of the entire population. The respondents were subsequently asked to log in with their Facebook account, and if willing, the respondent's public Facebook ID was stored. 2) Data was independently collected from all public profiles of Danish parties and politicians on Facebook. This includes posts, post-likes, comments, comment-likes, replies, reply-likes and tags. 3) After completion of step 1 and 2, we linked each respondent to the collected Facebook data, applying LASSO based multinomial logistic regression model to compare predict voter intention from Facebook data.*

# Results

The results depict how different measures of 'political likes' are able to predict which of the 9 parties in the Danish parliament a given person would vote for. The significance of the results is held against a Null-hypothesis that denotes no relationship between present day voter intention and explanatory variables. ,($H_0$: P = 1/9).

**Establishing a baseline from sociodemographic, political values and opinions**

We initiate our analysis establishing a baseline model based on sociodemographic variables, political values and opinions toward current issues collected through survey questions. The questions were selected to mirror the most typical variables for explaining voter alignment within the discipline of political science (20). The baseline model (model 0) includes 19 different features though many coefficients are neutralized by the L1 regularization. The optimal model makes predictions with 35.8% accuracy (CI of 2.9%) including 101 out of 668 coefficients. This closely mirrors the accuracies of similar questionnaire studies within political science, on average reaching an accuracy of approximately 35% (e.g. 20–22). For comparison reasons, we calculate the same model's accuracy for predicting present day voting intention on a right vs. left scale. Not surprisingly the accuracy is much higher when using this binary classification (80.1% accuracy).

**The power of a single political like**

With an established baseline model, we turn toward our collected Facebook data. As an initial experiment, we create a model that uses just a single feature, the latest like that the respondent has made to a post by a party or politician. This very simple setup (model 1) is more accurate and, on average, marginally better than our baseline model. With an accuracy of 43.9% (CI ±4.4%) and a right/left accuracy of 81.3%, model 1 indicates that a person's single latest political post-like tend to say more about party choice than a prediction model trained on a sample with 19 different features on each person, including questions on core political values.

**Raising accuracy by including peoples' entire political like history**

We now include all political likes for each person collected during the two-year period (model 2). The features in this model consist of the number of posts that a person has liked for each of the 9 parties in parliament. For example, if a respondent has liked a post made by a party, or a politician from that party, on their public Facebook page, then that counts as one like to that party for that respondent. To compare respondents who are extremely active on public pages with those who are less active, all values are normalized across. Applying these features, we predict which party a person would vote for with an accuracy of 60.9% (CI ±3.1%). This result is notably better than both the Baseline Model and Model 1. Interestingly, the best L1-penalty in Model 2 was 0.0 meaning that excluding coefficients would not increase the cross-validated accuracy. With a right/left average accuracy climbing to 90.1% the model suggests political likes as an efficient predictor for voter intention.

**Combining survey and political likes minutely increases prediction rate**

We now consider the possibility of positive complimentary effect by combining the best from two worlds. We add in the features of the baseline model to model 2 in order to explore if the survey questions drawn from political science literature are encapsulating other dimensions than the political likes: Do the two approaches overlap or complement each other? The new model, model 3, hereby includes all the sociodemographic background information, core political values from the baseline model, and the entire political like history from Model 2. The prediction accuracy is now 62,0% with (CI ±3.9%). This is greater than Model 2, but still within margin of error with 95% confidence. The increase in AUC and in right/left accuracy, however suggest that the model itself is slightly better than model 2.

Why is the combination of the two data sources not raising prediction rates more? One probable reason is the ratio between number of features and sample size. In Model 3 the sample size is 659, but has 27 features producing a total of 680 coefficients, which is greater than the sample size. Even with L1 regularization, which causes only some of the coefficients to be included (300/680) at optimal L1-penalty (8.0), the extra information added from Model 2 to Model 3 might only serve as noise. Thus, it is conceivable that the LASSO algorithm if employed on a bigger dataset would have found a somewhat better pattern, hereby raising the effective accuracy.

**Optimizing political likes prediction rates with minimum like criteria**

The previous models all propose political post likes as the single strongest variable for predicting individual party choice. It is therefore reasonable to consider if we can further optimize the use of this variable. Since we normalize the values for number of posts liked across each of the 9 parties for each respondent, our models might make overconfident predictions based on respondents who have only liked a single political post. Also, it is sensible to assume that a person, for whom 90% of her likes goes to posts from the same party, might be more likely to vote for that particular party compared to someone who have liked posts from 3 or 4 parties an equal number of times. In figure 1.1, we explore the relationship between these two criteria, namely 1) *Minimum likes,* excluding respondents with less total likes than the threshold and 2) *Party Like Cap,* excluding respondents with less percentage of likes directed toward a single party than the threshold. Because the two criteria involve filtering out respondents effectively cutting down the sample size as shown in Table 1.0, it is unfeasible to rely on the training of machine learning algorithms for classification. Thus, we made a simple algorithm, which makes predictions based on the party a respondent has liked the most at different intersections of the two criteria as shown in figure 1.1. We see that the simple model with Min Likes = 1 and Party Like Cap = 0.0 has an approximate accuracy of 56%, only slightly lower than Model 2, which used the same exact criteria but optimized by machine learning. Figure 1.1 shows how the accuracy increases with higher Min Likes and higher Party Like Cap meaning that, overall, the more likes a respondent has made as well as the percentage of likes that a respondent has made to a single party both increase the accuracy of the prediction. With e.g. Min Likes = 7 and Party Like Cap = 0.8 prediction accuracy goes above 90%, however sample size is down to 153, which also increases the error considerably.

**Thresholding total likes greatly increases accuracy**

Most importantly figure 1.1 shows that thresholding individuals on their total likes begin to converge significantly with a total minimum of 7 political likes. Setting the minimum likes criterion higher than 7 would gain only little in total accuracy, but significantly reduce sample size. We therefore interpret a threshold of political likes on 7 as the best choice for a near optimal prediction rate.

Based on the optimization exploration we deploy a fourth and final model that has the same features as Model 2, but only includes respondents with a total of 7 or more likes made to posts from parties or politicians. The effective
sample size is now 468 and average prediction accuracy has increased to around 70% with a CI of 5.1%. It is indicative of better prediction rates by imposing a criterion for how many political likes a user should have in total. Accuracy for right/left is now 96%.

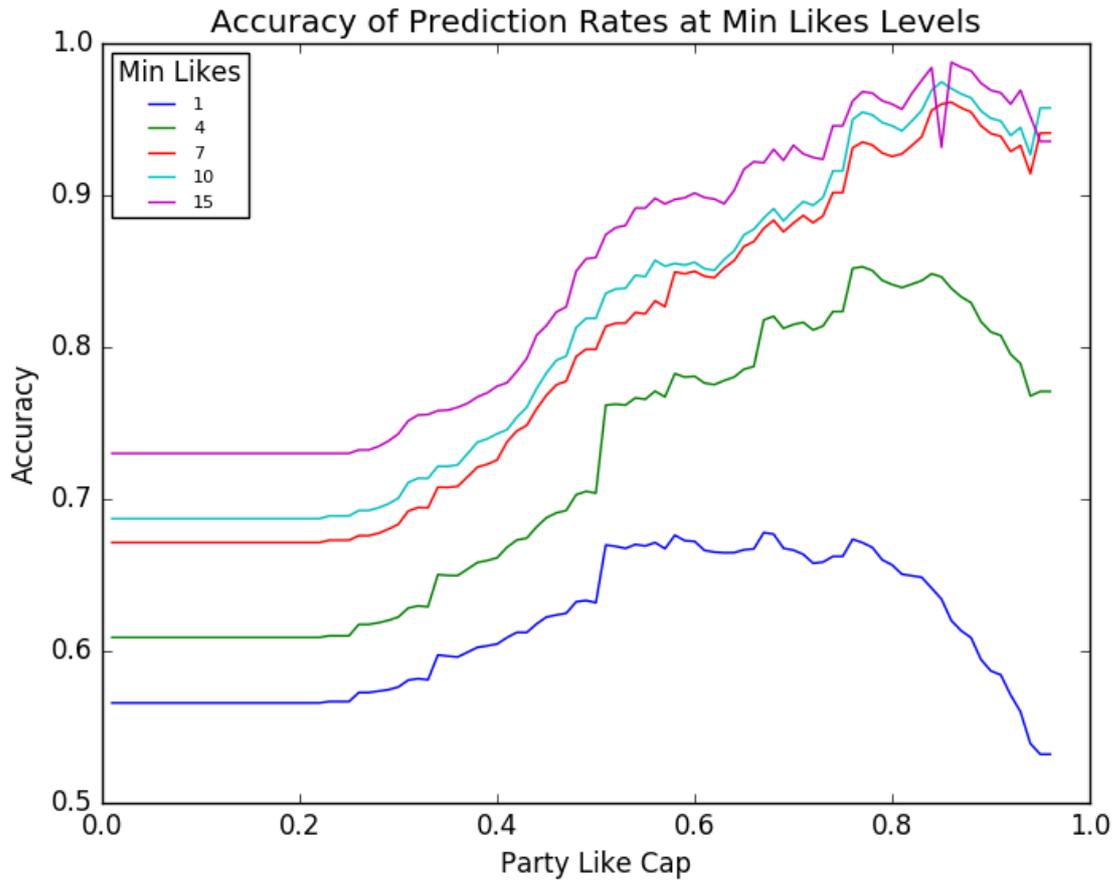

*Figure 1.1: The X-axis shows Party Like Cap (PLC) which denote how many likes in percentage that, at least, goes toward only a single party, meaning that at PLC = 0.8 only users who have at least 80% likes toward a single party are included. The Y-axis shows the percentage of users who are accurately labeled.*

| Party Like Cap | 0.0 | 0.5 | 0.7 | 0.8 | 0.9 |
|---|---|---|---|---|---|
| Sample size | 468 | 328 | 197 | 153 | 97 |
| 95% Confidence interval | 0.046 | 0.05 | 0.058 | 0.059 | 0.062 |
| Accuracy | 0.64 | 0.777 | 0.861 | 0.912 | 0.93 |

Table 1.0: Prediction rates and sample sizes at different party like caps with Min Likes 7 ($P < 0.001$).

# Table 1.1: Multinominal logistic regression models predicting present day voting

|  | **Baseline Model*** | **Model I *** | **Model II*** | **Model III*** | **Model IV*** |
|---|---|---|---|---|---|
| *Description* | Sociodemographic political values, current issues | Single latest political like† | All political likes | Model II + Baseline model | All political likes with Min Likes 7 |
| *Sample size* | 659 | 659 | 659 | 659 | 468 |
| *L1-Penalty* | 10.0 | 0.0 | 0.0 | 3.0 | 0.0 |
| *Incl./excl. Coefficients* | 101/668 | 90/90 | 90/90 | 300/680 | 90/90 |
| *± 95% Confidence Interval‡* | 0.029 | 0.044 | 0.031 | 0.039 | 0.051 |
| *Precision* | 0.461 | 0.420 | 0.652 | 0.567 | 0.755 |
| *Recall* | 0.302 | 0.449 | 0.577 | 0.591 | 0.612 |
| ***Accuracy*** | **0.358** | **0.439** | **0.609** | **0.620** | **0.708** |
| *Left/Right Acc.* | 0.803 | 0.813 | 0.908 | 0.933 | 0.96 |
| ***AUC*** | **0.760** | **0.762** | **0.832** | **0.878** | **0.917** |

*** = p < 0,001

† Latest political like refers to the like that a given person has made at the point in time closest to where she filled out the questionnaire.
‡ The confidence interval is calculated for the cross-validation error.

## Implications and limitations

A main implication of our results is the potential for studying political behaviour in multiparty systems on social media on a large scale and in near real time. The profiling of individual users through their political like history thus lends itself as a tool to study political participation on the social medias. Through collecting political likes, we become able to profile approximately 1.3 million Danes - 23% of the entire population - with a least one political like and 1 million with at least seven. By filtering posts based on political segments defined by millions of likes, the approach offers scholars, practitioners and politicians a view into the democratic voter's political dreams and the issues they engage in - all in a scale hitherto unknown in the discipline of political science.

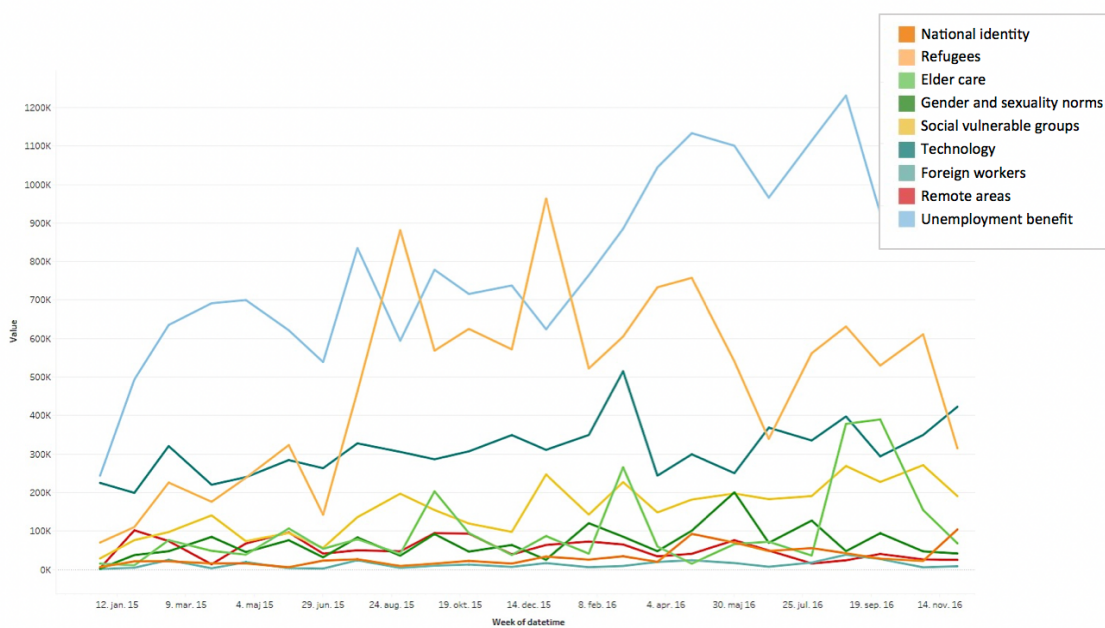

*Figure 1.2: Example showing trending topics for a group of users aligned with a specific party through their likes.*

The technique is limited to the 23% of Danes who have liked political actors within the studies 2 years' timeframe. Generalizing our findings outside this group requires attention towards the skewness inherent to the Facebook platform, most prominently toward the younger generations, but also with a slight bias toward women[§]. Further, one should also expect Danes with political likes to be slightly more political active than the remaining population (23). Our own future studies, i.e. predicting the aggregated electoral outcome based solely on political likes, are however comparable to most opinion polls, suggesting that our results can potentially be generalized to represent the entire Danish population (details in S7 Election prediction). Further, tentative explorations of neighboring multiparty countries (Sweden, Germany, Norway) show no indications that the amount nor the usage of political likes should vary radically across national borders. On this background and in accordance with studies made on other social media platforms (13), we estimate that the general mechanism should be reproducible in most open western multiparty democracies where Facebook is used as a political arena.

---

[§] Table S4 held against Facebook audience insights, https://www.facebook.com/ads/audience-insights/people?act=41292822&age=18-&country=DK.

**Toward parsimonious data**

Searching for patterns within big datasets of overly unrelated traces is not limited to Konsinski and colleagueas' work. Rather, the trend toward bigger and broader datasets seem to have become the standard for data experiments within the field of computational social science (e.g. 24). Increasingly, the ideal has also made it into commercial data analysis as illustrated when Cambridge Analytica, a data analytics firm drawing on (1)'s work, proclaimed to have secured Donald Trump his victory through the collection of "4-5,000 data points on every American" (25).

We share our colleagues from computational social science fascinating for the many stories about human behavior that apparently unrelated data traces can tell us. However, we also argue that the field of computational social science by now have reached a level of maturity that makes it timely to replace the ideal of *broad* data with a parsimoniousness ideal of *selective* data. While the studies by Kosinski and colleagues should not be compared 1:1 (due to differences in goals and context) it seems fair to note how using only the respondents single latest political like delivers a performance comparable to their best results built on hundreds of likes (1, 2)[**]. Future studies will reveal which other areas this strategy can be applied to, however, if our preference for 'Hello Kitty' and 'Harley Davidson' can reveal our personal traits, then what bigger expectations should we not hold towards the future of predictions built upon selective data linked to traits of interest.

---

[**] AUC. 0.8 vs. 0.85. (26) also reach a result of AUC = 0.8 using the same dataset as (1). We are here reporting the left-right AUC of our study in order to make our results comparable.

## Why liking predicts voting? Contours of a theory

*Accuracy* with *generalizability* is thus the main advantage of our parsimonious data strategy. Based solely on this limited data scope consisting of the single latest like per respondent, we were able to predict multiparty choice with an accuracy of 0.439. The accuracy could be lifted above 0.6 by including all likes and above 0.7 by imposing a minimum like criteria of 7 likes. Our results thus indicate that even a single political like is comparable in accuracy to most studies in political science, commonly reaching an accuracy of around 35% through combining survey question on sociodemographic, political values and opinions toward current issues (e.g. 20–22). While this line of research is not entirely comparable (the political scientists searching for explanation rather than prediction), the predictive power of political likes becomes striking when contemplating the approximately 30 survey questions involved in reaching these 35%.

On this background, it seems reasonable to consider why likes predict voting behaviour dramatically better than survey questions; what makes likes predictive of our political behaviour in the first place. Referencing main theories within studies of voting behaviour, one could suggest that like is predictive because it shows ideologically alignment with the party (3), that the issue taken up by a politician is a current political issue relevant to us (27, 28), or that we share an admiration for the personal traits of the candidate (29, 30).

Our response to this question is to re-articulate our initial claim: that likes comprise a generic mechanism for users to show their support. Political likes should be seen as a measure that captures a multitude of the above-mentioned - and probably also other - theories for *why* we vote (i.e. ideology, shared issue or personal identification). This response is in line with the overall high accuracies reached, which makes it difficult to imagine one single theoretical driver, along with the lack of complimentary effects seen in model 3 suggest that we should view likes as encapsulating a majority of different motives and preferences.

The high accuracies and the lack of complimentary effects also indicate that most people are highly selective with their political likes. We should thus not think of political likes as a cost-free interaction that we carelessly direct toward any post that catches our attention, but rather as an interaction form that we apply when we are clearly aligned across one or even multiple axes of preferences. As such, political like should be seen as a behavioural measure that condensates a heterogeneous mixture of different motives and individual's inscription into politics.

**Materials and methods**

**Data:** We base our prediction on likes made on page posts by Danish parties and politicians collected from January 2015 to 2017 through the public Facebook Graph API. Likes are a generic mechanism used by Facebook users to express their support of content that have already shown as a good proxy for predicting both electoral results and personal traits (1, 2, 10), making it an immediate choice for exploring the possibility for predicting political orientation. For use in the final analysis, we only include respondents, who were able, and willing, to share their public Facebook ID (N = 1216). We also limit our analysis to respondents who had liked political actors during the period and would vote for any of the 9 parties currently in parliament (N = 659). The final sample is slightly smaller (~23%) than what we theoretically would expect given our database of political likes containing data from 1,4 million Danes. We ascribe most of this dropout to privacy concerns (See also Table S3). As a result of this dropout, representativeness of the data sample becomes slightly distorted (See non-response analysis in S3). However, for the most parts, the distortion simply reproduces Facebook's already skewed user groups with the only large skew being an underrepresentation of older users; a skew which recently have been shown to have limited affect effect on how often a person would like political actors (23).

**Method:** We test post like against a baseline model developed from questionnaire data on socio-demographics, political values and opinions on political issues developed based on current 'best-practice' within political science (See Table S5). Going forth from the baseline model, we gradually compare a selection of multinomial logistic regression models all predicting which party a person would vote for, but modelled on different selections of Facebook data as well as combinations of Facebook and questionnaire data. Using L1 regularization LASSO (31), only features that contribute significantly to the overall prediction are included in the models. In each model an L1-penalty was selected using 10-fold cross validation to avoid overfitting and account for variance in the prediction accuracy. The process is illustrated in Figure 1.0 while the models and results can be found in table 1.1.

**Acknowledgments.**
We thank Sune Lehmann, Samuel Roberts, Anders Blok, Michael Bossetta, Vedran Sekera, Piotr Sapiezynski, Enys Mones, and Snorre Raulund for their critical reading of the manuscript. T.B was supported by a grant from the KU16 funding, J.B.K from University of Canterbury, UC Doctoral Scholarship. We also want to acknowledge Nextwork A/S and Analyse & Tal F.M.B.A, for allowing T.A., E.D, M.J. and M.S to participate in the process during work hour and for lending us access to their database.

# SUPPLEMENTARY INFORMATION

PARSIMONIOUS DATA: HOW A SINGLE FACEBOOK LIKE PREDICTS VOTING BEHAVIOUR IN MULTIPARTY SYSTEMS

## Table of Content

# S1. DATA COLLECTION

## *QUESTIONNAIRE DATA*

We collected data from a random sample of 3050 individuals through a European survey company. At the beginning of the questionnaire the respondents were given the option to connect their answers to their public Facebook profile. Interviewees were instructed that by agreeing to the terms we would get access to their public Facebook ID, but not any of their private content. As illustrated in the table below the main sample of 3050 individuals vary from the national population on key variables by a maximum of 4 percentage points.

*Table S1: Distributions of users in our questionnaire compared to the population statistic. East – West indicates which side of the country the respondent lives in.*

| Quotas | % | Population | Dif |
|---|---|---|---|
| **Male-18-34 Years of age -East** | 5% | 7% | -2% |
| **Male-18-34 Years of age -West** | 6% | 8% | -2% |
| **Male-35-53 Years of age -East** | 9% | 7% | 2% |
| **Male-35-53 Years of age -West** | 11% | 8% | 3% |
| **Male-54-74 Years of age -East** | 6% | 9% | -3% |
| **Male-54-74 Years of age -West** | 9% | 11% | -2% |
| **Female-18-34 Years of age -East,** | 11% | 7% | 4% |
| **Female-18-34 Years of age - West,** | 10% | 8% | 2% |
| **Female-35-53 Years of age -East,** | 7% | 7% | 0% |
| **Female-35-53 Years of age - West,** | 10% | 8% | 2% |
| **Female-54-74 Years of age -East,** | 6% | 9% | -3% |
| **Female-54-74 Years of age - West,** | 10% | 11% | -2% |
| | **100%** | **100%** | |

All data collected from interviewees used in this study were done with their full consent with each respondent retaining the right to have their data deleted at any time.

*FACEBOOK DATA*

All data were collected using the public Facebook Graph API 2.7[1]. The data covers the period from January 2015 and January 2017 and harvested from a combination of media pages and political pages. The data has been collected from 378 political pages covering politicians in parliament (or persons who have previously run for parliament) along with pages for political parties. The media pages total 116 and consist of the most popular print and online papers in the country along with a selection of other popular media like blogs and magazines. The exact amount of data points in the database to this date is shown in the table below.

*Table S2 the full Facebook Dataset collected.*

|  | *political pages (n=378)* | *media pages (n=116)* |
|---|---:|---:|
| **posts** | 345.967 | 637.349 |
| **post-likes[2]** | 78.117.089 | 141.351.908 |

## S2. COMBINING FACEBOOK AND QUESTIONNAIRE DATA

In order to use our two data collections (questionnaire and Facebook) we had to create a sample cross cutting both collections. To explore different filters and samples we used a progressive filtering procedure meaning we kept adding filters gradually. The exploration is illustrated in the table below. The filter represented in each row includes both the filter described and all the above filters.

---

[1] https://developers.facebook.com/docs/graph-api/reference/v2.7/.

[2] The introduction of emojis as an alternative to the generic likes overlaps with our data collection period. We include all emojis under the same category: 'likes'.

*Table S3 Data filtering*

| Filter | Sample size |
|---|---|
| No filter, all respondents in questionnaire | 3050 |
| Only respondents who reported their public Facebook ID in questionnaire | 1216 |
| Only respondents who had liked at least one post on political pages corresponding to a party in parliament *(used in baseline and model II - IV)* | 659 |
| Only respondents who had liked at least 7 posts on political pages *(used in model V)* | 468 |

We want to compare all our prediction models with the baseline model. Thus, we use the filtered sample (N = 659) in case the particular sample is overall easier to predict than the full questionnaire for the baseline model and models 1 – 4 we use the sample that includes all respondents with a political like. For model 5 we use only users with at least 7 likes (N = 468). Potential skews and biases as a product of the filtering process is covered in the non-response analysis.

### S3. NON-RESPONSE ANALYSIS

In order to address the significant decrease in sample size from the full questionnaire to the samples being used in our models, we have conducted a non-response analysis. The primary reason for the drop in respondents is people who did not have a Facebook profile or wouldn't allow access to their public ID (N = 3050 -> 1216). Accounting for majority of this drop are the 40% of the Danish population who do not have a Facebook account. Also, we expect privacy concerns to withhold a number of users from participating. The second drop is when we exclude respondents who would not vote for one of the 9 parties or have not made at least one political like during the period (N = 1216 -> 659). Accounting for the majority of this drop are people who do not plan to vote or who is not between the 28% (1,3 million) of the Danes who have liked political actors and therefore not present in our database.

*Table S4 Results of Chi squared permutation tests*

|  | Total survey (N = 3050) | | | With Facebook ID (N= 1216) | | | With political likes (N = 659) | | | Degree of Skew | |
|---|---|---|---|---|---|---|---|---|---|---|---|
| Category | X-squared mean | 0.025 quantile | 0.975 quantile | X-squared mean | 0.025 quantile | 0.975 quantile | X-squared mean | 0.025 quantile | 0.975 quantile | N = 1216 | N = 659 |
| Party choice | 8.08 | 3.11 | 15.23 | 17.42 | 9.81 | 27.67 | 19.91 | 9.92 | 29.24 | Small | Small |
| Individual responsibility vs. Public responsibility | 2.78 | 0.32 | 7.55 | 6.69 | 2.03 | 13.38 | 14.89 | 6.83 | 24.62 | Not significant | Medium |
| Losing entitlement vs. Right to choose job | 2.80 | 0.30 | 7.92 | 7.11 | 2.35 | 13.17 | 7.92 | 2.70 | 15.42 | Not significant | Not significant |
| Social security reforms have become too many vs. Just enough | 2.80 | 0.38 | 7.51 | 3.38 | 0.67 | 7.61 | 6.08 | 1.86 | 12.94 | Not significant | Not significant |
| Competition is healthy vs. Unhealthy | 2.83 | 0.34 | 8.03 | 6.84 | 2.08 | 13.25 | 3.76 | 0.62 | 8.76 | Not significant | Not significant |
| More freedom to corporation vs. Less freedom | 2.87 | 0.31 | 7.85 | 1.95 | 0.23 | 5.05 | 5.90 | 1.53 | 12.61 | Not significant | Not significant |
| People with high salary do not pay enough tax | 2.81 | 0.34 | 7.33 | 4.21 | 0.85 | 9.20 | 7.00 | 2.06 | 14.19 | Not significant | Not significant |
| Income inequality is too high | 2.69 | 0.37 | 7.36 | 3.62 | 0.72 | 8.33 | 4.08 | 0.96 | 9.12 | Not significant | Not significant |
| Violent crimes should have be punished harder | 2.79 | 0.32 | 7.37 | 1.89 | 0.22 | 5.17 | 2.63 | 0.38 | 7.23 | Not significant | Not significant |
| More border control | 2.85 | 0.38 | 7.80 | 6.97 | 2.21 | 13.87 | 7.83 | 2.50 | 15.09 | Not significant | Not significant |
| We should do more to protect national heritage | 2.82 | 0.33 | 7.81 | 5.81 | 1.61 | 11.62 | 6.63 | 2.12 | 13.67 | Not significant | Not significant |
| We should prevent crime through counseling rather than punishment | 2.79 | 0.38 | 7.79 | 6.06 | 1.71 | 12.27 | 6.84 | 2.02 | 13.55 | Not significant | Not significant |
| Environment vs. Corporate expansion | 2.93 | 0.44 | 8.04 | 4.57 | 1.17 | 9.89 | 8.34 | 2.83 | 15.48 | Not significant | Small |
| Homosexuals should have exactly the same rights as everyone else | 2.71 | 0.32 | 7.70 | 7.95 | 2.84 | 14.19 | 8.32 | 2.94 | 15.62 | Small | Small |
| More tax on gasoline | 2.91 | 0.37 | 7.82 | 9.44 | 3.80 | 17.21 | 7.38 | 2.46 | 14.30 | Small | Not significant |
| Religious extremist have the right to public gatherings | 2.82 | 0.38 | 7.58 | 8.49 | 3.02 | 16.02 | 7.65 | 2.44 | 15.36 | Small | Small |
| Gender | 0.65 | 0 | 3.52 | 0.97 | 0 | 3.78 | 4.49 | 0.71 | 10.46 | Not significant | Small |
| Age | 1.33 | 0.03 | 4.59 | 27.09 | 16.69 | 39.90 | 14.59 | 6.75 | 25.87 | Large | Medium |
| Geography | 2.72 | 0.30 | 7.42 | 3.29 | 0.63 | 7.46 | 4.05 | 0.84 | 9.18 | Not significant | Not significant |
| Education | 5.53 | 1.46 | 12.54 | 16.98 | 6.62 | 28.92 | 24.74 | 14.84 | 36.18 | Small | Small |

In order to find out how skewed the questionnaire features were, we did a 10000-fold permutation test of Chi-squared scores for each of the samples (see Table S4). The first test compares the distributions of answers from two random samples both from the full questionnaire (Total survey N = 3050). The next compares random samples where one is from the full questionnaire and the other is from a sample of only respondents with attached Facebook profiles (On Facebook N = 1216). The last compares random samples where one is from the full questionnaire and the other is from a sample of only respondents with at least one political like on any of the parties in parliament (With political like N = 659). The distribution of answers corresponding to a single feature e.g. age is only considered to be skewed if the mean of the X-squared value, resulting from the permutation test, lies outside of the 95% confidence interval of the first permutation test that compared two random samples from the full questionnaire. For example, in the N = 1216 sample, Gender has an X-squared mean of 0.96, but the 95% confidence interval of N = 3050 has X-squared values between 0 and 3.5, so the skew is not significant. However, the N = 659 has a Gender X-squared mean of 4.49 and is therefore considered to have a small skew; Whether a skew is small or large is determined by the relative skew from the largest to the smallest. It is important to remember that the degree of skew for a single feature is determined by its relation to the *statistical* significance and not how skewed it actually is. For example, the most skewed feature, age, has a 10-percentage points difference between young and old.

Based on this exploration we see how the only large skew is age, with an underrepresentation of older users. A recent study in the neighbour country of Norway, however found that age had very limited effect on how often a person would like political actors (22). The same study also found that women and people with lower education were more prominent in liking political actors, which are in line with our dataset's smaller skews along these characteristics (See non-response analysis in S3). While the representativeness is thus within the expected bounds of a study on Facebooks, we should remain attentive to how neither groups of people who do not act publicly or are not on Facebook at all are not absent in the study (23).

## S4. POST LIKES NORMALIZATION

Post-likes are the simplest and most useful feature of all our Facebook features for predicting party choice. A vector represents each user with each field in the vector corresponding to the number of times that person has liked a post from a given party or politician from that party. As shown in the figure below, number of post likes per user has a big head long tail distribution.

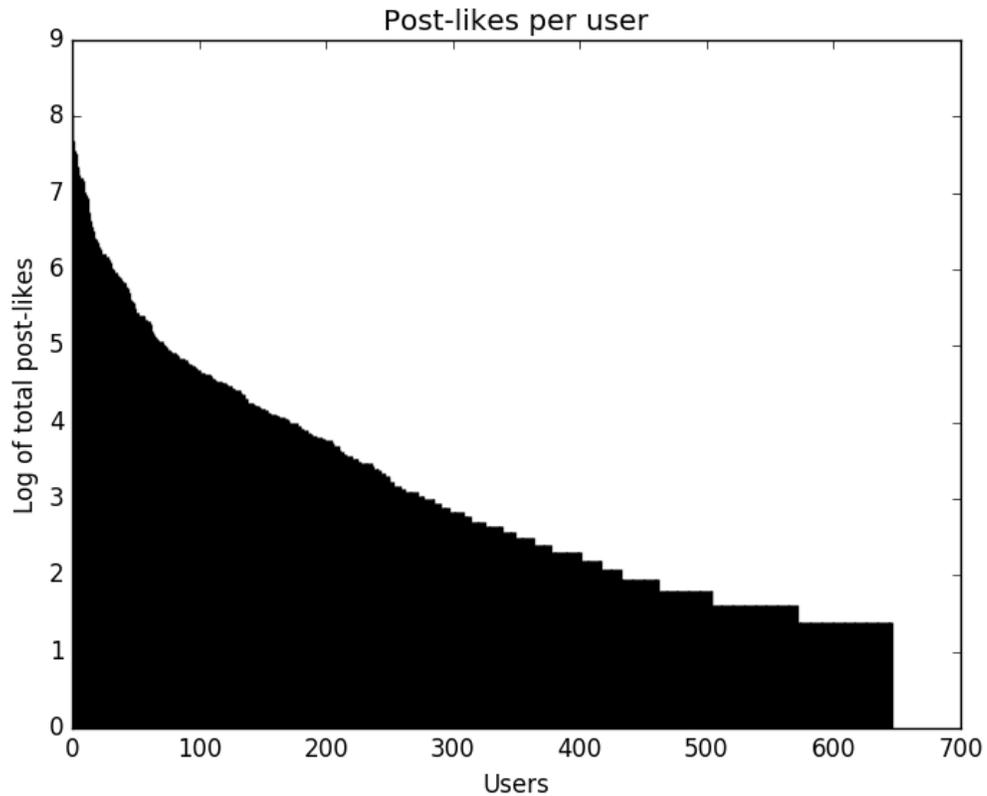

*Figure S1 Log of post likes per user in the main sample used in the models (N = 659).*

For this reason, the number of times that each user has liked posts from a party is divided by the total number of posts the user has liked. This effectively normalizes the sum of the vector for all users, which makes it possible to compare very active users with less active ones. An illustration of the creation of the normalized post-likes vector is shown below.

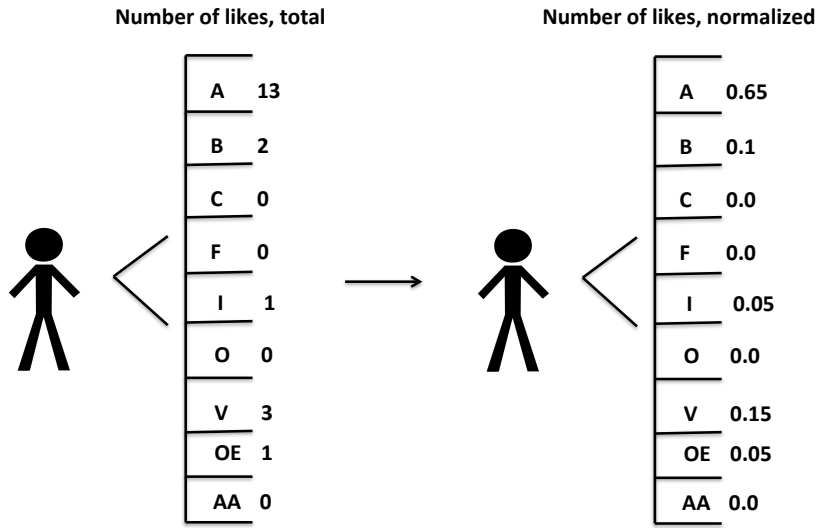

*Figure S2 Users' post likes normalization process.*

## S5. INCLUDING BEHAVIOUR OF FRIENDS AND LIKEMINDED USERS

The assumption behind tags is that they are primarily being used between friends since you have to know the name of the person you are tagging, even though it can also be used to tag adversaries. This assumption combined with the preliminary results showing that users who like posts from a party are likely to vote for that party is the basis of the value calculated for tags. So, a user gets one count for a certain party depending on what party that the tagging/tagged user has liked.

Example:

- User A has the post-likes vector [0, 0, 0, 0.4, 0, 0.4, 0.2, 0, 0]
- User B has the post-likes vector [0, 0, 0, 0.9, 0.1, 0, 0, 0, 0]

If user A tags user B, then:

- User A has tags vector [0, 0, 0, 0.9, 0.1, 0, 0, 0, 0]
- User B has tags vector [0, 0, 0, 0.4, 0, 0.4, 0.2, 0, 0]

The process is cumulative, so a new tags vector is added to the previous for each user for each tag that a user has made or received.

The same procedure is used for comment-likes, however only comments from politicians' pages or from media posts that have a high probability of having political content are included.

High probability of political content builds on the idea that posts with political content create political conjecture. It is therefore determined by the cosine distance between the aggregated post-likes vectors of users who have liked a post on a media page and the aggregated vectors of users who have commented on the same post. Only the 25% of media post with the highest cosine distance are considered to be high probability of political content.

To validate this, a random sample of 300 posts each were drawn, respectively, from the 25% of media post with the highest cosine distance and the remaining 75% of media posts. Posts were manually labeled as either political or non-political and the two samples were compared to each other. Posts from the sample with high cosine distances contained almost exclusively political posts while the other sample was more of a mix. A Chi Squared test was performed to validate the statistical significance ($p < 0.0001$).

## S6. REGRESSION MODELS

All data analysis is made with Python. The code used for the regression models can be attributed mainly to Turi's GraphLab Create[3]; all other data analysis is original code. The data models are all based on GraphLab Create's Logistic Regression module and implement only L1 regularization (L2 is set to 0). L1 regularization is used for selecting the coefficients corresponding to the features, which deliver the best bias-variance tradeoff for generalizing the model. L1 regularization performs this selection by setting least important coefficients to exactly zero while decreasing other coefficients by a value relative to the chosen λ-value. The least important coefficients can roughly be defined as those least related to (least correlated with) the maximum log likelihood (MLE) of the model under a certain λ-value. The relation between MLE and a given coefficient is determined by soft thresholding(1, 2).

---

[3] https://turi.com/products/create/docs/

We report the following evaluative measures, all in the form of cross-validated averages:

> ***AUC*** (Area Under the ROC-curve): Effectively states the overall ability of the regression model for separating classes based on input variables with 0.5 denoting no relationship between explanatory variables and prediction rate. The thresholds for the ROC-curve are incremented by 0.0001.
>
> **Precision:** The number of true positives out of all positives, TP/(TP+FP). For multiclass calculated as the mean of all classes.
>
> **Recall:** The number of true positives out of all positives and false negatives, TP/(TP+FN). For multiclass calculated as the mean of all classes.
>
> **Accuracy:** The global amount of correctly predicted classes divided by the total sample size of the test data. In multiclass cases this is not the mean of all class accuracies.

TABLE S5 BASELINE MODEL – THE PREDICTIVE STRENGTH OF INDIVIDUAL FEATURES

| Category | AUC | Sample Size |
|---|---|---|
| Gender** | 0.617 | 659 |
| Age** | 0.610 | 659 |
| Geography | 0.525 | 659 |
| Education** | 0.624 | 659 |
| Individual responsibility vs. Public responsibility *** | 0.712 | 659 |
| Losing entitlement vs. Right to choose job*** | 0.653 | 659 |
| Social security reforms have become too many vs. Just enough** | 0.639 | 659 |
| Competition is healthy vs. Unhealthy*** | 0.655 | 659 |
| More freedom to corporation vs. Less freedom*** | 0.687 | 659 |
| People with high salary do not pay enough tax*** | 0.700 | 659 |
| Income inequality is too high*** | 0.686 | 659 |
| Violent crimes should have be punished harder | 0.585 | 659 |
| More border control*** | 0.687 | 659 |
| We should do more to protect national heritage*** | 0.686 | 659 |
| We should prevent crime through consueling rather than punishment** | 0.609 | 659 |
| Enviroment vs. Corporate expansion*** | 0.687 | 659 |
| Homosexuals should have exactly the same rights as everyone else | 0.497 | 659 |
| More tax on gasoline** | 0.639 | 659 |
| Religious extremist have the right to public gatherings** | 0.603 | 659 |

*** = p < 0,001     ** = p < 0,05

## S7. ELECTION PREDICTION

To test the generalizability of our findings we explore the ability for raw likes to predict the Danish general election in 2015. Our initial attempt work by counting each person as one vote toward the party that she has liked the most. Through this approach, we reach a Mean Absolute Error (MAE) of 0.0271 not far from the MAE of polls from the same period reaching averaging on 0.010. Upon closer examination, it however becomes apparent that the main cause of error is linked to an overrepresentation of the Red-Green Alliance and an underrepresentation of the social democrats. This is in line with the age-skew in our data toward younger participants, with the red-green are known for their great shares of younger voters where the social democrats have an overrepresentation of older voters.

On this premise, we develop an approximate weighting scheme that takes two random opinion polls produced at least a year before election and use them to create weights for the Facebook counts. Weight optimization process is a simple machine learning process that minimizes the Residual Sum of Squares (RSS) between the Facebook count and the opinion polls chosen as the input. The Facebook count is made for the week leading up to the date of the opinion poll. Through this procedure, MAE is decreased to 0.011.

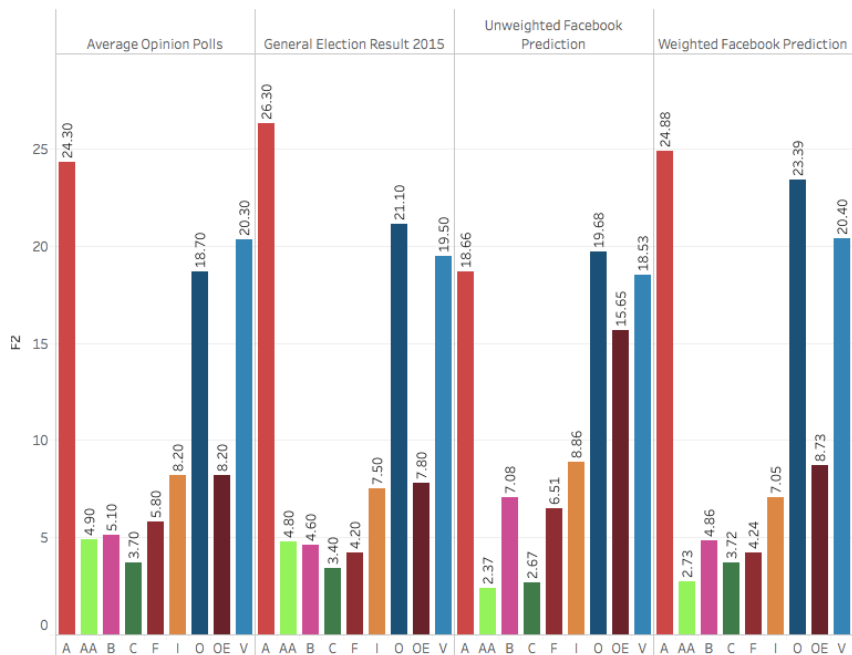

*Figure S3: Three predictions and the actual election results for the 2015 national election in Denmark.*